\documentclass{iopart}
\usepackage{iopams}
\usepackage{mathrsfs}
\usepackage[usenames,dvipsnames]{xcolor}
\usepackage[pdftex]{graphicx}

\newcommand{\bra}[1]{\left\langle #1 \right | }
\newcommand{\ket}[1]{\left | #1 \right\rangle }

\newcommand{\DeltaS}{{\Delta_{\mathrm S}}}
\newcommand{\DeltaT}{{\Delta_{\mathrm T}}}
\begin{document}

\title[Draft version]{Probing Majorana fermions in the tunneling spectra of
a resonant level}
\author{R. Koryt\'ar$^1$ and P. Schmitteckert$^{1,2}$}
\address{$^1$Institute for Nanotechnology, Karlsruhe Institute of
Technology (KIT), 76021 Karlsruhe, Germany}
\address{$^2$ DFG Center for Functional Nanostructures, Karlsruhe Institute of Technology (KIT), 76128 Karlsruhe, Germany}
\ead{richard.korytar@kit.edu}
%==============================================================================
\begin{abstract}
Unambiguous identification of Majorana physics presents an outstanding
problem whose solution could render topological quantum computing feasible.
We develop a numerical approach to treat finite-size superconducting
chains supporting Majorana fermions, which is based on iterative application
of a two-site Bogoliubov transformation. We demonstrate the applicability
of the method by studying a resonant level attached to the superconductor 
subject to external perturbations. In the topological phase, we show that the 
spectrum of a single 
resonant level allows to distinguish peak coming from Majorana physics from 
the Kondo resonance.
\end{abstract}
\pacs{73.21.Hb,74.55.+v,74.78.Na}
%\keywords{}
\maketitle
\section{Introduction}
Condensed matter theory, in the course of its continued quiet revolution,
has endowed our understanding of nature by introducing quasi-particles
with diverse, often exotic, behavior.
A prominent example is a Majorana fermion\cite{majorana:nc},
a particle being its own
antiparticle. The possibility of synthesizing Majorana fermions has
 attracted a lot of attention,
because of their potential for quantum computing. Information carried
this way would be essentially non-local and retrieved by certain non-Abelian
operations (braiding), being rather immune to general disturbances of
the environment\cite{wilczek:nphys,beenakker,alicea:np,alicea:rpp}. Although several condensed matter
environments could support these zero
modes\cite{zoller:prl,readgreen:prb,zhang:prb,ivanov:prl,Thomale_Rachel_PS:X2013},
topological superconductors \cite{fukane:prl,kitaev:uspeki,Leijnse:sst}
have emerged as a natural playground.
The latter are not restricted to rare materials, rather they are
engineered easily by forming appropriate 
heterostructures with ordinary $s$-wave 
superconductors\cite{dassarma:prl,vonoppen:prl}.
Experimental setups, which, based on predictions, could host Majorana 
fermions,
have been prepared\cite{mourik:science,shtrikman:nphys,furdyna:nphys,xu:nl},
but their unambiguous experimental observation
is a task far from being obvious.
A very promising route is offered by transport measurements, since
Majorana modes should give rise to a zero-bias resonance which does
not react to weak changes of magnetic field or gate voltage.

However, conductivity enhancement near zero bias is a common
companion of diverse collective phenomena. An example is Kondo effect, 
0.7 anomaly \cite{cronenwelt:prl,rokhinson:prb} or recently proposed
electronic disorder \cite{bagrets:prl}. Thus, a careful elimination
of these scenarios should be a part of the ``smoking - gun'' probe
of the Majorana particle.
We would like to contribute to this debate by studying
a superconducting system whose local spectral function exhibits
resonances coming from two sources: the topologically protected
Majorana state and a single-particle electronic resonance.

Our model system consists of a single level weakly coupled to a one-dimensional
superconductor. We consider both singlet and triplet pairing mechanisms.
We firstly verify, that the resonance of the single level survives after coupling to
a singlet-paired superconductor. We observe that the position of the
resulting resonance is coupled to the external gate voltage.
When singlet pairing is replaced by the triplet one, the superconductor 
model is the well
known Kitaev chain \cite{kitaev:uspeki} doubled due to spin degeneracy,
having Majorana states at both ends of the
chain\footnote{Per spin, there is a single fermionic boundary mode, which
corresponds to two separate Majorana states}.
These modes give rise to resonances in the local spectrum at zero energy.
The nature of the coupling between one of the Majorana states to the resonant 
level (RL) is revealed in the behavior of the resonances upon local gating 
and weak magnetic field. Among the plethora of peaks, the RL
manifests itself due to the coupling to the external fields, in contrast
to the Majorana peak. Thus, the distinctive behavior of the peak structure
allows us to suggest a new means of experimental demonstration of
Majorana physics.

In order to see the behavior of the resonances, it is sufficient to look
at local spectral functions, which directly convey the electron/hole 
tunneling spectral density.

The calculation of spectra involves solving for the eigenstates of the
Hamiltonian with mean-field superconducting fluctuations, i.e. a general
Hermitian operator which is bilinear in fermionic operators.
We employ a numerical diagonalization method based on iterative application
of Bogoliubov transformation \cite{bogoljubov} to a Fock subspace spanned by two orbitals.
This is an extension of the Jacobi diagonalization procedure:
not only we consider basis change, but also general
automorphisms of the Fermi operator algebra.

From a theoretical standpoint, this approach relies on a fully 
fermionic language. Thus, it is capable at providing intuitively transparent
arguments, like occupation numbers, in contrast to an alternative approach, 
based on Majorana representation of the Hamiltonian \cite{Wimmer2012algorithm}. 
From the experimental point, results of our work suggest a means to identify 
the character of close-to zero-bias resonances, and distinguish them
from peaks originating from different sources, such as the charge and spin
resonances of the low-temperature Anderson model.

Our paper is structured as follows: in the next section we introduce
the model Hamiltonian and the diagonalization procedure. The two-orbital
``dimer'' not only serves as an illustration of the method, but presents
a single iteration step when large chains are treated. At the end of
section \ref{Sec:modelmethods} we detail on the calculation of the
local spectral function. In the Results section, we show the spectra
calculated for a RL coupled to a singlet-paired and triplet-paired
superconductor, contrast the differences and  understand the trends of the 
peaks. At the end of this work, we discuss the experimental relevance of our
calculations and summarize.

\section{\label{Sec:modelmethods}Model and methods}
\subsection{Hamiltonian}
We study a resonant level (RL) coupled to a one dimensional spin-full superconducting chain.
The Hamiltonian of the full system reads
\begin{eqnarray}
\nonumber
H = \epsilon^{\phantom\dagger}_{0\sigma}  \hat c^\dagger_{0\sigma} \hat c^{\phantom\dagger}_{0\sigma} + 
t_0\left( \hat c^\dagger_{0\sigma} \hat c^{\phantom\dagger}_{1\sigma} + \mathrm{h.c.}\right) \\
\nonumber
- \mu_\sigma\sum_{i=1}^N  \hat c^\dagger_{i\sigma} \hat c^{}_{i\sigma} 
- t\sum_{i=1}^{N-1}\left( \hat c^\dagger_{i+1\sigma} \hat c^{\phantom\dagger}_{i\sigma} + \mathrm{h.c.}\right)\\
+\DeltaS\sum_{i=1}^N\left( \hat c^\dagger_{i\uparrow} \hat c^{\dagger}_{i\downarrow} + \mathrm{h.c.}\right)
+\DeltaT\sum_{i=1}^{N-1}\left( \hat c^\dagger_{i+1\sigma} \hat c^{\dagger}_{i\sigma} + \mathrm{h.c.}\right) \,,
\label{Eq:rlham}
\end{eqnarray}
where $\hat c^{}_i$ ($\hat c^{}_i$) are fermionic anihilation (creation) operatoers at site $i$.
The orbital labeled 0 is the RL coupled weakly to the rest of the chain,
which spans the sites $1,...N$. The first and second term are RL's
on-site energy and hopping
energy to the chain, respectively. The second line contains the chemical potential 
$\mu_\sigma$ of 
the chain (spin-dependence masks a Zeeman field)
and the hopping term between nearest neighbour pairs of the chain.
The last two terms describe superconducting paring in the mean-field fashion: the singlet
pairing with strength $\DeltaS$ and triplet pairing proportional to $\DeltaT$.
Spin index $\sigma$, wherever appears in \eref{Eq:rlham}, implies summation.
\subsection{Diagonalization}
We solve for the eigen-energies and eigenvectors of the Hamiltonian \eref{Eq:rlham}
numerically. Before we explain the full procedure, we illustrate few important points
in the two-orbital case.
\subsubsection{\label{Sec:Methods:Dimer}Dimer}
We start with a simple two-orbital (dimer) Hamiltonian described by
\begin{equation} \label{Eq:H_Dimer}
H = \epsilon \hat c^\dagger_1 \hat c^{\phantom\dagger}_1 +
    \epsilon' \hat c^\dagger_2 \hat c^{\phantom\dagger}_2 +
    t\left(\hat c^\dagger_2 \hat c^{\phantom\dagger}_1 + 
    \hat c^\dagger_1 \hat c^{\phantom\dagger}_2\right) +
    \Delta\left(
    \hat c^{\phantom\dagger}_2 \hat c^{\phantom\dagger}_1 + 
    \hat c^{\dagger}_1 \hat c^{\dagger}_2\right)
\end{equation}
with two un-equal on-site terms, a hopping term and the last
term describes superconducting pairing.
For convenience, the Hamiltonian is written in a matrix form
\begin{equation}
\label{Eq:ham}
H = \left(\hat c_1^\dagger\ \hat c_2^\dagger\ \hat c_1\ \hat c_2\right)
 \left(\begin{array}{cccc} 
                \epsilon & t & 0 & \frac{\Delta}{2}\\
                 t & \epsilon' & -\frac{\Delta}{2} & 0\\
                 0 & -\frac{\Delta}{2} & 0 & 0 \\
                 \frac{\Delta}{2} & 0 & 0 & 0
 \end{array} \right)
 \left(\begin{array}{c} \hat c_1 \\ \hat c_2\\ \hat c_1^\dagger \\ \hat c_2^\dagger \end{array}\right).
\end{equation}
where the row and column vectors are similar to Nambu spinors, see~\ref{sec:NambuSpinors}.
We bring it to a diagonal form in a two-step procedure. Firstly,
we transform away the anomalous pairing terms by a Bogoliubov transformation.
Then the Hamiltonian attains a coupled two-level structure which
can be brought to a diagonal form by a standard unitary transformation
(ie a change of basis). 

We wish to subject the fermion operators to a linear transformation
\begin{equation}
\label{Eq:auto}
\left(\begin{array}{c} \hat d_1 \\ \hat d_2\\ \hat d_1^\dagger \\ \hat d_2^\dagger \end{array}\right)
= \mathscr U
\left(\begin{array}{c} \hat c_1 \\ \hat c_2\\ \hat c_1^\dagger \\ \hat c_2^\dagger \end{array}\right)
\end{equation}
and ensure that the new operators $\left\{\hat d^{\phantom\dagger}_i,\hat d_i^\dagger\right\}$
obey anti-commutation relations. It follows that the 
$4\times 4$ matrix $\mathscr U$ must be unitary.
We write it in the block form
\begin{equation}
\mathscr U = \left(\begin{array}{cc} U & V\\ V' & U'\end{array}\right)
\end{equation}
with each sub-matrix $U,U',V,V'$ containing $2\times 2$ elements,
so that $\hat d_i = U_{ij}\hat c_j + V_{ij}\hat c_j^\dagger$ and
$\hat d^\dagger_i=V'_{ij}\hat c_j + U'_{ij}\hat c_j^\dagger$ (Einstein summation).
It follows that $U'=U^*$ and $V'=V^*$ and the transformation matrix
must have the following structure
\begin{equation}
\label{Eq:structure}
\mathscr U = \left(\begin{array}{cc} U & V\\ V^* & U^*\end{array}\right).
\end{equation}
Unitarity implies
\begin{eqnarray}
\nonumber
UU^\dagger + VV^\dagger &= 1\\
\label{Eq:unitarity:uv}
UV^\top + VU^\top &= 0 \,,
\end{eqnarray}
with $V^\top $ ($U^\top$) the transposed of $V$ ($U$).
We detail in Appendix on how to choose $U$ and $V$ so that the
transformed Hamiltonian is free from anomalous terms of the form
$d_1d_2$.

Then, the Hamiltonian can be rearranged to a normal-ordered form
\begin{equation}
\label{Eq:ham2}
H = \mathbf{\Psi^\dagger}
 \left(\begin{array}{cccc} 
     \epsilon_1' & t' & 0 & 0 \\
      t' & \epsilon_2' & 0 & 0\\
      0 & 0 & 0 & 0 \\
      0 & 0 & 0 & 0
 \end{array} \right)\mathbf\Psi
\end{equation}
where we introduced $\mathbf\Psi^\dagger = \left(d_1^\dagger\ d_2^\dagger\ d_1\ d_2\right)$
We achieve the diagonalization by the transformation of the kind
(\ref{Eq:auto}) with the structure of the unitary matrix
\begin{equation}
\mathscr U'' =
\left(\begin{array}{cc}
U'' & 0\\ 0 & U''
\end{array}\right).
\end{equation}
This is, of course, a standard eigenvalue problem of a two-level system.

\subsubsection{Chain}
Our numerical diagonalization of the full Hamiltonian consists of iterative
application of the procedure shown in the preceeding section. Firstly, one
singles out two orbitals in the Hilbert space, then applies the Bogoliubov
transformation that removes pairing of the selected ``dimer'', then transforms
the remaining parts of the Hamiltonian accordingly.

Here we detail on the outlined procedure. Let us consider a general Hamiltonian 
of a Bogoliubov-de Gennes type
\begin{equation}
H = \mathbf{\Gamma^\dagger}\mathscr H \mathbf{\Gamma}
\end{equation}
where $\mathbf{\Gamma^\dagger} := \left(c^\dagger_1,\dots,c^\dagger_N,
c^{\phantom\dagger}_1,\dots,c^{\phantom\dagger}_N\right)$ is a row vector of
fermion operators labeled by indices of an orthonormal basis set 
(spin and orbital) and $\mathscr H$
is a $2N\times 2N$ matrix. The assignment of the matrix elements of $\mathscr H$
is ambiguous and in what follows we will require the following $N\times N$ block
structure
\begin{equation}
\label{Eq:hamform}
\mathscr H = \left(\begin{array}{cc} \phantom{-}h & b \\
 -b & 0 \end{array} \right)
\end{equation}
with $h^\dagger = h$ and $b^\dagger = -b$.
 The fermionic operator algebra allows for linear
automorphisms $\mathbf\Gamma' = \mathscr U \mathbf\Gamma$ by the means of a unitary
matrix $\mathscr U$ with the $N\times N$ block structure
\begin{equation}
\label{Eq:structure2}
\mathscr U = \left(\begin{array}{cc} U & V \\
 V^* & U^* \end{array} \right).
\end{equation}
We wish to find a matrix $\mathscr U$, which carries away all pairing and hopping
terms from the Hamiltonian. In addition we demand that the transformed operators
still obey fermionic anti-commutations relations which again leads to the condition
\eref{Eq:unitarity:uv}. It is important to note that this implies $\mathscr U$ being
unitary. However, an arbitrary unitary $\mathscr U$ matrix will in general not adhere to
the form \eref{Eq:structure2} and therefore not conserve the fermionic anti-commutation
relations. Therefore, just diagonalizing matrix ${\mathscr H}$ will in general not lead 
to a valid solution. One may argue that any unitary matrix that diagonalizes 
$\mathscr H$ should lead to the same solution. However, this does not apply here.
The transformation we describe below  will in general not lead to a completely diagonal matrix.
Instead it will transform the matrix into a form where the off-diagonal symmetric part of $h$ 
and the asymmetric part of $b$ vanishes. In addition the '0' in \eref{Eq:hamform} may not be zero.
Therefore, the resulting matrix only leads to a Hamiltonian that is equivalent to a diagonal form 
for fermionic operators.

In order to ensure the structure \eref{Eq:structure2}
we developed an iterative procedure, sketched
in the following steps (i-iii). Suppose that at the $p$'th step we were given
a matrix $\mathscr H_p$ -- of the form \eref{Eq:hamform}.
\begin{enumerate}
\item We identify a pair of orbitals with the largest pairing, say
      $m< n \le N$. The operator set
      $c_m,c_n,c^\dagger_m,c^\dagger_n$ is transformed to a new
      operator set $c'_m,c'_n,c'^\dagger_m,c'^\dagger_n,\ m< n \le N$ in which
      the pairing vanishes. This amounts to solving the ``dimer'' problem from the
      previous section.
\item The matrix $\mathscr H_{p}$ becomes 
      $\mathscr H_{p+1} = \mathscr U_p\mathscr H_p\mathscr U_p^\dagger$ when
      expressed in the new operator set, the matrix $\mathscr U_p$ does not mix
      operators with index other than $m$ or $n$; in the dimer Fock subspace
      has the structure as in equations (\ref{Eq:unitarity:uv},\ref{Eq:structure}).
\item The Hermitian matrix $\mathscr H_{p+1}$ is to be brought to a form 
      \eref{Eq:hamform} by a permutation of fermionic operators. Note, that this
      step eventually generates a $c-$number term to the
      Hamiltonian which can be discarded.
\end{enumerate}
After all pairings are lower than a certain tolerance, 
we perform a basis change to rotate away hoppings. This can be achieved
by the transformation matrix \eref{Eq:structure2} with zero off-diagonal
blocks and is a routine task.
All results presented in this work were calculated by this
iterative method; the Hamiltonian was thought converged if the absolute value
of the highest pairing was smaller than $10^{-10}t$. We collect the product
of all transformation matrices in order to calculate local quantities.

\subsection{Spectral function}
Let the Hamiltonian (\ref{Eq:rlham}) be transformed to a diagonal form
\begin{equation*}
H=\sum_i E_i \hat d_i^\dagger \hat d_i
\end{equation*}
by $d_i = \sum (U_{ij} \hat c^{\phantom\dagger}_j + V_{ij} \hat c^\dagger_j)$.

%The retarded Green's function is defined by
%\begin{equation}
%G_{ii}(t) = -i\bra{} \left\{c_i(t),c_i^\dagger(0)\right\}\ket{} \theta(t).
%\end{equation}
%This is an ''on-site'' component. Moreover, it is a normal-state
%component. The Fourier transform is
The normal-component retarded Green's function at site $i$ reads
\begin{equation*}
G_{ii}(\omega) = \bra{} \hat c_i\frac{1}{\omega-H+i\eta} \hat c^\dagger_i
+ \hat c^\dagger_i\frac{1}{\omega+H+i\eta} \hat c_i\ket{}.
\end{equation*}
%We insert the transformation (\ref{Eq:auto})
%\begin{eqnarray}
%\fl\sum_j\Biggl( U_{ij}^2\bra{} d_jd^\dagger_j\ket{} \frac{1}{\omega-E_j+i\eta}
%+ V_{ij}^2\bra{} d^\dagger_jd_j\ket{} \frac{1}{\omega-E_j +i\eta} +\\
%+U_{ij}^2\bra{} d^\dagger_jd_j\ket{} \frac{1}{\omega+E_j+i\eta} +
%V_{ij}^2\bra{} d_jd^\dagger_j\ket{}\frac{1}{\omega+E_j+i\eta}\Biggr).
%\end{eqnarray}
%We assume that $\bra{} d_jd_i\ket{}=0$ while 
%$\bra{} d^\dagger_jd_i\ket{} = f(E_j)\delta_{ij}$, the Fermi-Dirac
%distribution function.
The spectral function is given by the expression
\begin{equation}
A_{ii}(\omega) = -\frac{1}{\pi}\Im\sum_j\left[ 
\frac{\left|U_{ij}\right|^2}{\omega-E_j+i\eta} + \frac{\left|V_{ij}\right|^2}{\omega+E_j+i\eta}\right]
\end{equation}
where $\eta$ is a positive
infinitesimal, which in the numerics will be substituted by a finite
value of the order of the average level splitting around the Fermi
level in the absence of superconductivity. Hereafter we shall use
$\eta=0.015 t$. The resulting spectra are deconvoluted by the algorithm
of the reference \cite{Peter:deconvl} (poor man's deconvolution).
\section{Results}
From now on, all numerical values of dimension energy (such as pairings, 
on-site energies, \emph{etc}.) will be in the units of $t$.

%\subsection{Spectral properties of a resonant level}
The Hamiltonian \eref{Eq:rlham} is studied in two regimes:
\begin{enumerate}
\item a resonant level coupled to a singlet-paired superconductor
      ($\DeltaT$ = 0,$\DeltaS$ = 0.6)
\item a resonant level coupled to a triplet-paired superconductor
      ($\DeltaS$ = 0,$\DeltaT$ = 0.3)
\end{enumerate}
Thus, the superconducting bulk gap will be $2.4$ in both cases. The
rest of the chain Hamiltonian is parameterized as $\mu=0$ and $t=1$ as the
unit of energy. In all cases, the superconducting chain has 200 sites.
The RL is weakly coupled to the chain with
$t_0 = 0.3$ and on-site energy $\epsilon_{0\sigma}$ will vary.

Hence, the case (ii) is in a topologically non-trivial phase \cite{kitaev:uspeki} with
Majorana modes. The first case is a topologically trivial superconducting
state.

\subsection{Behavior of local spectral function on gating}
\begin{figure}
\includegraphics[width=\columnwidth]{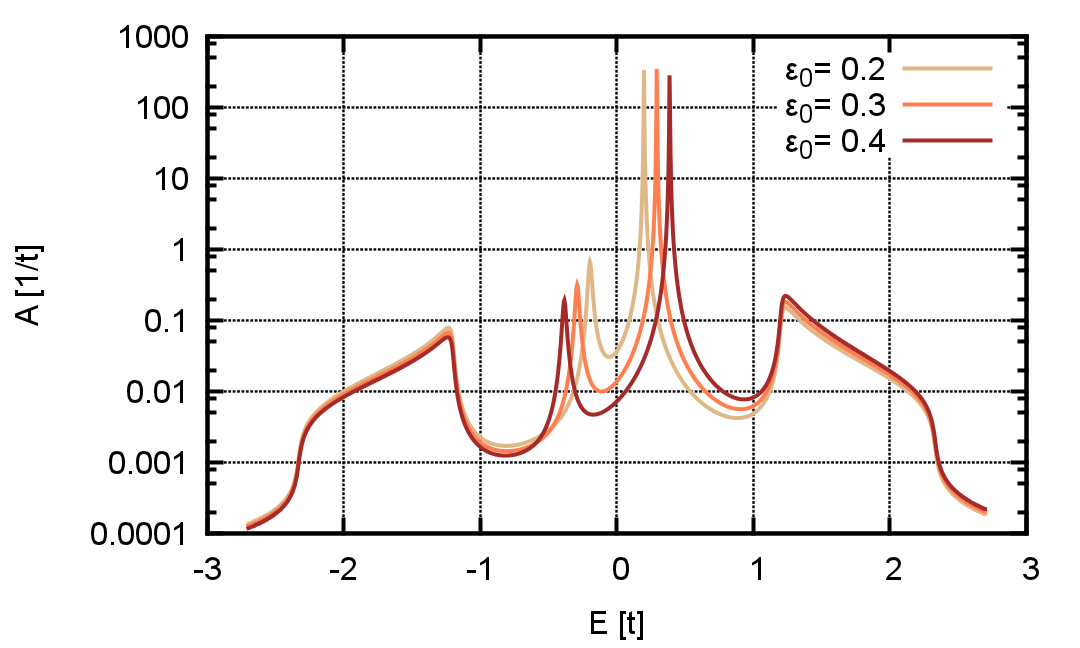}
\caption{\label{Fig:onsite-s}Spectral function of a resonant level coupled to
a singlet-paired superconductor chain with 200 sites. The resonant level couples via hopping
$t_0=0.3$ to the chain. Three curves correspond to different on-site energies
of the level.}
\end{figure}
\begin{figure}
\includegraphics[width=\columnwidth]{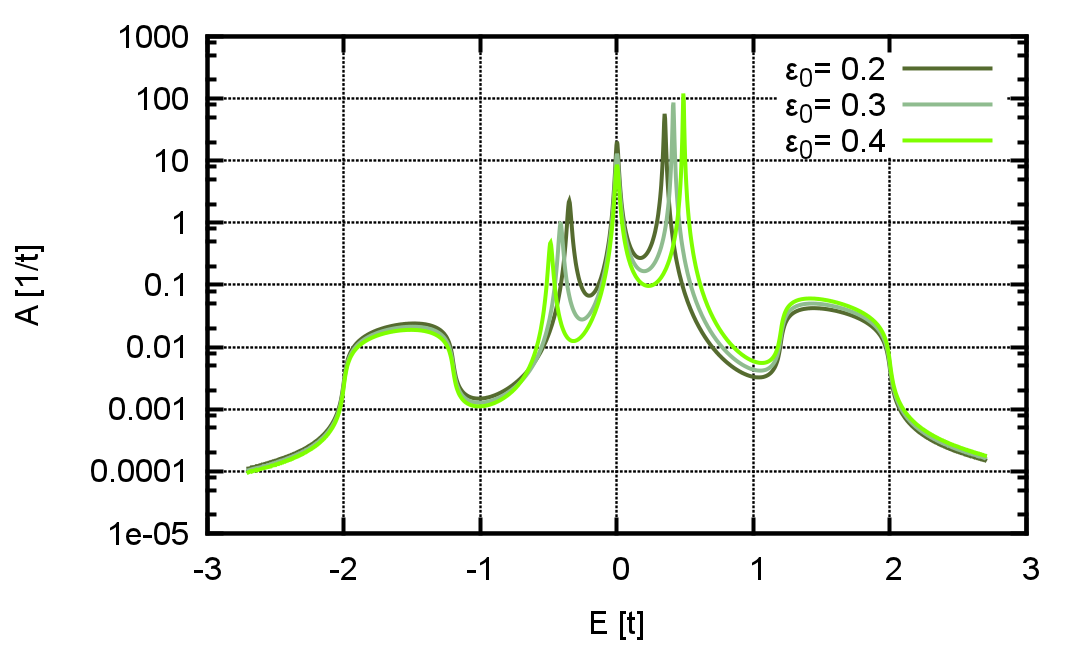}
\caption{\label{Fig:onsite-t}Spectral function of a resonant level coupled to
a triplet-paired superconductor chain with 200 sites. The resonant level couples via hopping
$t_0=0.3$ to the chain. Three curves correspond to different on-site energies
of the level.}
\end{figure}
In figure~\ref{Fig:onsite-s} we show several spectral function for case (i).
There are no zero-modes. The position of the highest peak coincides with the 
on-site energy of the resonant level $\epsilon_0$. Along with this RL
peak, there is an Andreev-reflected peak roughly at $-\epsilon_0$. The 
mirroring behavior of the resonant and Andreev peak positions as the
``gate'' varies is a known distinctive feature of the Andreev process.
Finally, in the outside region $|E| > t$ two ridge-like features
appear. These are the bulk features of an ordinary superconductor.

Now we proceed to the case (ii), the topological superconductor, 
figure~\ref{Fig:onsite-t}. A strikingly new feature is the zero-energy peak,
which signals the spreading of the Majorana mode to the RL site.
Note that the position of the RL is
repelled from the bare value, the resonance center is shifted upwards.
This can be understood as a delocalization of the
Majorana fermion: without the resonant state, the Hamiltonian in regime (ii)
reduces to a spin-full Kitaev chain with zero modes localized at both edges.
Coupling to the RL dilutes the Majorana state 
\cite{MarcoPolini:prb}; however, since
it's energy is fixed to zero, the coupling-induced splitting can affect the
RL position only.

\subsection{Magnetic field dependence of local spectral function}
After seeing the distinct behavior of peaks under gating, we proceed to study
the dependence of spectra in a weak magnetic field applied to the
whole system. We introduce
a homogeneous Zeeman field by substituting $\epsilon_{0\sigma}
= \epsilon_{0} + \sigma B/2$ and $\mu_\sigma = \mu + \sigma B/2$, $\sigma=\pm 1$
in the Hamiltonian \eref{Eq:rlham}. Hence, $B$ is the spin splitting
of a single level.

\begin{figure}
\includegraphics[width=\columnwidth]{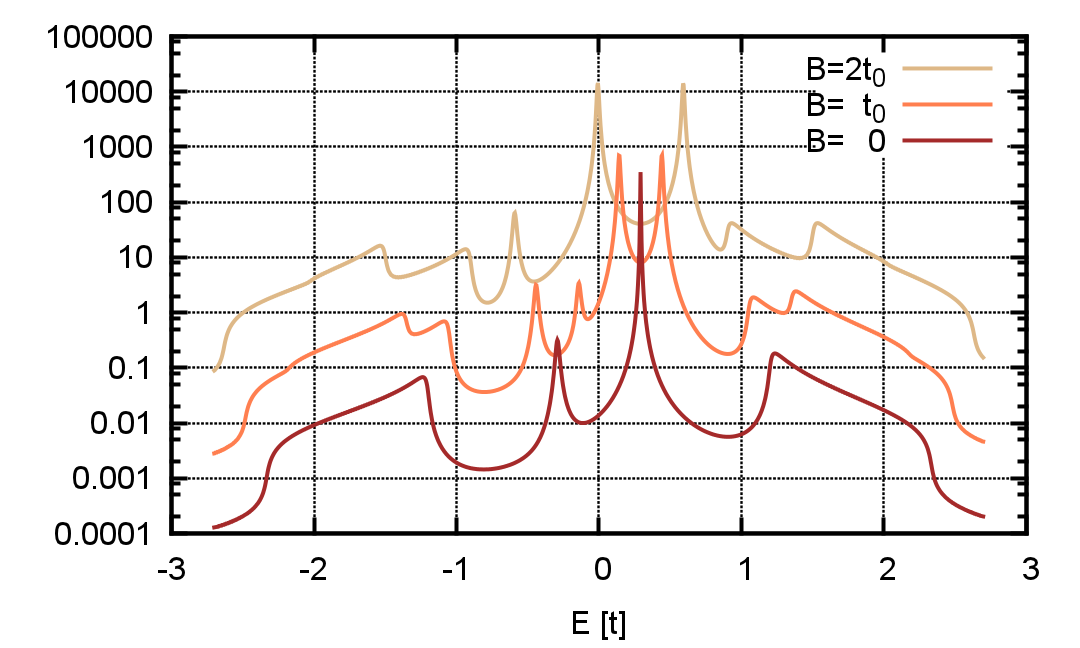}
\caption{\label{Fig:zeeman-s} Spectral function of a resonant level weakly coupled to
a singlet-paired superconductor with 200 sites. The three curves represent different
values of a homogeneous Zeeman field, each curve is a sum over two projections of spin.
$B$ denotes the splitting of a single level
and is expressed in multiples of the resonant level hopping $t_0$. For clarity,
the spectra have been shifted vertically.}
\end{figure}

In the singlet paired case the RL and it's Andreev reflected
counterpart split, as shown in figure~\eref{Fig:zeeman-s}.
Note that in the highest splitting $B=2t_0$ an \emph{accidental} zero-energy
resonance forms from the overlap of two resonances.
The magnitude of peak splitting obeys strictly the Zeeman term in the RL Hamiltonian.

As the last case, we treat the influence of Zeeman field on the
triplet-paired superconductor, case (ii)\footnote{We remark that this part
applies to the realizations of topological superconductivity where spin
degeneracy remains.}.
The pinning of the Majorana state to the zero energy (the chemical 
potential) is robust against weak magnetic field, too. The magnetic
field, however, splits the resonant and Andreev levels.
\begin{figure}
\includegraphics[width=\columnwidth]{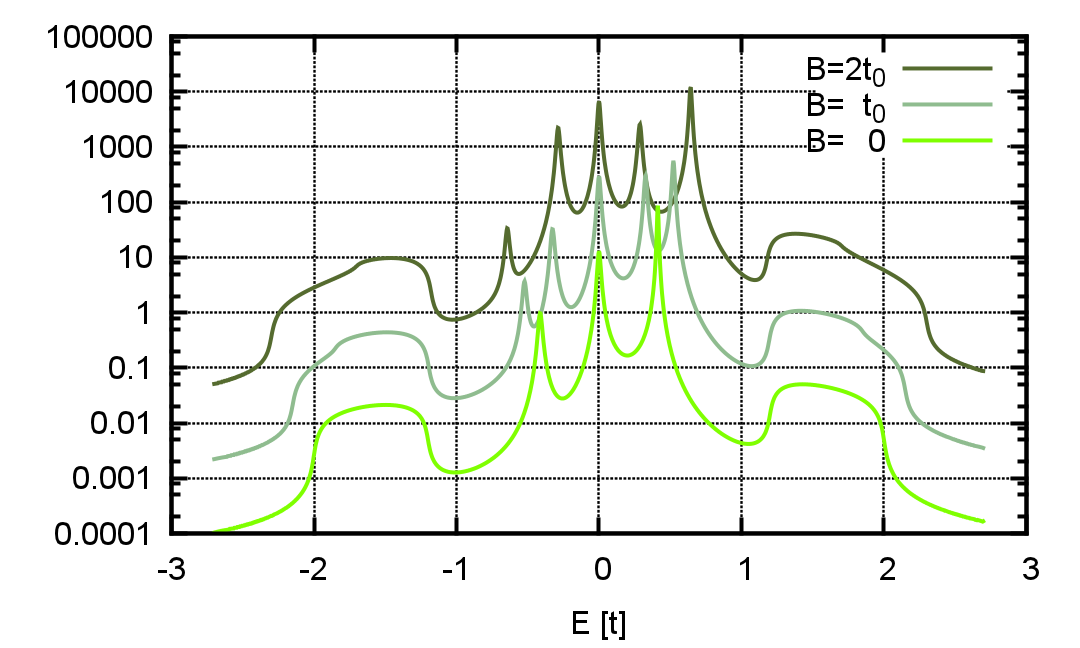}
\caption{\label{Fig:zeeman-t} Spectral function of a resonant level weakly coupled to
a triplet-paired superconductor with 200 sites. The three curves represent different
values of a homogeneous Zeeman field, each curve is a sum over two projections
of spin. $B$ denotes the splitting of a single level
and is expressed in multiples of the resonant level hopping $t_0$. For clarity,
the spectra have been shifted vertically.}
\end{figure}
Figure~\eref{Fig:zeeman-t} shows evolution of the spectral function
for the Zeeman fields $B=0,t_0$ and $2t_0$. The side peaks are clearly
split while the central resonance stays almost intact. Again, all peaks
``interact'' and apart from the previously mentioned shift of the Andreev and resonant
levels off the zero energy, there is also a decrease in the spin splitting below
the value given by $B$. 
\section{Discussion}
As stated in the introduction, zero bias transport features can arise
in solid-state systems due to reasons unrelated to Majorana physics.

A thoroughly studied example is Kondo effect. In the transport through
quantum wires, Kondo physics could emerge due to enhancement of Coulomb
interactions ( due to reduced screening) or presence of a magnetic impurity
in the wire. The spectral manifestation at low temperatures is the zero-energy
Kondo resonance (width of the order of the Kondo temperature $k_BT_K$) and 
possibly charge peaks of the Anderson model. The side peaks 
need not be symmetric around the Fermi level and behave differently
when external fields are introduced. A change of the gate voltage
(on-site energy of the Anderson impurity) causes the lower and higher side peaks move
\emph{in the same sense}, accompanied by a pronounced change in the width of the 
central (Kondo) resonance. This is in sharp contrast with the case of a RL coupled to a Majorana fermion (compare to figure \ref{Fig:onsite-t}),
where side peaks move off the zero bias symmetrically and the central resonance
stays intact.

Further distinction emerges when studying the Zeeman splitting. Majorana resonance
does not split even though the side peaks do (figure~\ref{Fig:zeeman-t}). 
The Kondo resonance, in contrary, is immune only for weak fields and splits
when the Zeeman field becomes of the order of $k_BT_K$, ie of the order of the
peak width. The resulting splitting of the zero energy peak is then twice the 
Zeeman splitting. Hence, magnetic field and local gating are both independently
sufficient to distinguish between both underlying mechanisms.

\section{Conclusions}
We have developed an efficient numerical approach capable to treat 
hybrid inhomogeneous systems. 
Our method relies on a fully fermionic formalism, that is, eigenstates
of the Hamiltonian are fermions.
We diagonalize the Hamiltonian by the means of a unitary matrix and
point out that the matrix structure must be restricted in order to satisfy 
anti-commutation algebra.

We have applied the method to the combined system of a resonant level
(RL)
and a superconducting chain. Regardless of the nature of pairing,
gate voltage and magnetic fields couple to the RL, as we
show in the analysis of the level's spectral function. 
When triplet pairing is introduced,
the local spectral function attains a zero-bias resonance,
which does not react to external fields. In summary, the combined
system allows to prove that the perturbing field does couple to the
system, at the same time demonstrating the presence of a Majorana 
mode.

Apart from a direct access to spectral functions,
the fermionic language could be conveniently used to calculate other
local quantities, as for instance wave functions of the zero modes.
Alicea \emph{et al.} \cite{alicea:np} have elaborated on the
possibility to implement quantum memory in a network of topological
superconducting wires. Here, the braiding operation could be realized by
changes in local gate voltages. Our method, straightforwardly 
extended to non-stationary regimes, offers a way to 
simulate braiding while tracking the Majorana state in real time.

\ack
PS would like to thank Stephan Rachel, Pascal Simon, Ronny Thomale, and Johannes Reuther
for stimulating discussions.
%====================================================================================
\appendix
\section{Single angle Bogoliubov transformation}
In section~\ref{Sec:Methods:Dimer} we introduced general conditions for
the transformation matrix of the dimer fermion operators.
In order to rotate the Hamiltonian in a particle conserving representation
we have to rotate the  asymmetric part of $B$ to zero.
in recipe described above we have reduced the problem to
solving a two site system which can be achieved via a standard fermionic Bogoliubov transformation
\begin{eqnarray}
 \hat{d}^{}_1 &=&  \cos(\beta) \hat{c}^{}_1 \,-\, \sin(\beta) \hat{c}^{+}_2 \\  
 \hat{d}^{}_2 &=&  \cos(\beta) \hat{c}^{}_2 \,+\, \sin(\beta) \hat{c}^{+}_1  \,,
\end{eqnarray}
which fulfill the conditions \eref{Eq:unitarity:uv}.
In matrix form they are given by
\begin{equation}
\label{Eq:SingleAngleUV}
       U= \left(\begin{array}{cc}  \cos(\beta) & 0 \\ 0 & \cos(\beta) \end{array} \right)
\qquad V= \left(\begin{array}{cc}  0 & -\sin(\beta)  \\  \sin(\beta)  & 0\end{array} \right) \,.
\end{equation}
Using
\begin{equation}
\tan(2\beta) = -\frac{2\Delta}{\epsilon+\epsilon'}.
\end{equation}
the $\Delta$ contribution to Hamiltonian $H$ in \eref{Eq:H_Dimer} is tranformed
to zero.
\section{Two angle Bogoliubov transformation}
In addition to transforming the $\Delta$ part to zero, we can also rotate the $t$ contribution
to zero by applying an additional rotation
\begin{equation}
\label{Eq:ToAngleUV}
       R= \left(\begin{array}{cc}  \cos(\alpha) & -\sin(\alpha) \\ \sin(\alpha) & \cos(\alpha) \end{array} \right)
\quad U_R= R \cdot U \quad V_R = R \cdot V \,,
\end{equation}
where the rotation angle is given by
\begin{equation}
\tan(2\alpha) = -\frac{2t}{\epsilon-\epsilon'}.
\end{equation}
\section{Remarks on Nambu spinors} \label{sec:NambuSpinors}
In the case of $s$-wave pairing one usually resorts to Nambu spinors
\begin{equation}
 \mathbf\Psi_x^\dagger = \left( \hat{c}_{x,\uparrow}^\dagger\ \hat{c}_{x,\downarrow}^\dagger\ \hat{c}^{}_{x,\downarrow}\ -\hat{c}^{}_{x,\uparrow} \right) \,.
\end{equation}
Using the label $j=2x$ for the up spins and $j=2x+1$ to the down spins this corresponds to a matrix representation
using a transformation of the form
\begin{equation}
 \mathscr U_{\mathrm N} = \left(\begin{array}{cc} 1 & 0\\ 0 & U_{\mathrm N} \end{array}\right) \,,
\end{equation}
where $U_{\mathrm N}$ block diagonal consisting of $2 \times 2$ rotation matrices with an angle of $\pi/2$.
Using this transformation one is led to a matrix where the new $\tilde{b}$ block is actually
symmetric. 
The disadvantage of the Nambu representation is that the conditions for preserving canonical
anti-commutation relations are not as simple as eq. \eref{Eq:unitarity:uv} in our representation.

%However, for this to happen the anormalous terms  have to have a special structure,
%like only $s$-wave BCS terms. In detail, restricitng to real Hamiltonians,
%one is looking fo a matrix $U_{\mathrm N}$ with the property
%\begin{equation}
%  \tilde{b}^+ = \tilde{b} \quad   \Rightarrow  \quad U_{\mathrm N} b^+ =  - U_{\mathrm N} b = b U^+_{\mathrm N} \,.
%\end{equation}
%Therefoe one needs transformation matrix with the property
%\begin{equation}
%     U_{\mathrm N} b U_{\mathrm N} = -b  \,,
%\end{equation}
%which does exist in general.
%%
\section{Remarks on the numerics}
For the recipe of bringing the many site problem to diagonal form it is
sufficient to use the single angle version.
In our tests it turned out that the two angle version actually
needs less iteration step to converge. However, each step costs approximately
twice as the number of vector operations is doubled.
In return the single angle version was typically faster.

One can also start the procedure by first rotating $b$ into a tridiagonal form 
using the algorithm of \cite{Wimmer2012algorithm}.
Moreover, an anti-symmetric matrix can be transofrmed into a $2\times 2$-block diagonal matrix $D = Q \cdot b \cdot Q^\dagger$ with $Q$ unitary \cite{Ward_Gray_ACM1978,Golub_Loan:MC}.
Since $D$ consists only of $2\times 2$ and $1\times 1$ blocks with a diagonal of zeroes  $D^2$ is diagonal.
Therefore we have
\begin{equation}
  D^2 = Q \cdot b \cdot Q^\dagger \cdot Q \cdot b \cdot Q^\dagger =  Q \cdot b^2 \cdot Q^\dagger  \,,
\end{equation}
where $Q$ is given by the diagonalization of $b^2$. Note, that it is essential that $b$ is used in the antisymmetric form $b = -b^\top$.
Starting our procedure with $U=Q$ and $V=0$ leads to an intial $2\times 2$-block diagonal  matrix $b$ leading to an improved convergence rate
in our tests.

\vspace*{2ex}

\bibliographystyle{unsrt}
\bibliography{references}
\end{document}